\documentclass{PoS}

\newcommand{\lsi}{LS~I~+61$^{\circ}$303}
\newcommand{\ls}{LS\,5039}
\newcommand{\hess}{HESS J0632+057}
\newcommand{\apjl}{ApJ}
\newcommand{\apj}{ApJ}
\newcommand{\aap}{A\&A}

\newcommand{\ergcms}{erg cm$^{-2}$ s$^{-1}$}
\newcommand{\psrb}{PSR B1259$-$63}

\usepackage{enumitem}

\title{High energy $\gamma$-ray emission from \psrb\ during 2014 and 2010 periastron passages}

\ShortTitle{\psrb\ periastron passages on 2014 and 2010}

\author{\speaker{Giuseppe Andrea Caliandro}\\
        W. W. Hansen Experimental Physics Laboratory, Kavli Institute for Particle Astrophysics and Cosmology, Department of Physics and SLAC National Accelerator Laboratory, Stanford University, US. 
        Consorzio Interuniversitario per la Fisica Spaziale (CIFS), Italy\\
        E-mail: \email{caliandr@slac.stanford.edu}}

\author{C. C. Teddy Cheung\\
Space Science Division, Naval Research Laboratory, USA}

\author{Jian Li\\
Institut de Ci\`encies de l'Espai (IEEC-CSIC), Spain}

\author{Diego F. Torres\\
Institut de Ci\`encies de l'Espai (IEEC-CSIC), Spain
Inst. Catalana de Recerca i Estudis Avancats, Spain}

\author{Kent Wood\\
Space Science Division, Naval Research Laboratory, USA}

\abstract{\psrb/SS 2883 is a $\gamma$-ray binary system composed of a radio pulsar in a long (1236.7 days) and elliptical ($e\sim0.87$) orbit around a Be star. 
In its 2010 periastron passage, multiwavelength emission from radio to TeV was observed, and an unexpected GeV flare was detected by the Fermi Large Area Telescope (LAT). 
Here we present the results of the LAT monitoring of \psrb\ during its most recent 2014 periastron passage. 
We confirm that the GeV flare is recurrent within the orbit. The comparison of the 2014 and 2010 periastron passages shows overall similarities of flare durations, average flux levels, and spectra. In contrast, the detailed time evolutions of the two flares present interesting differences. Indeed, the light curves of the two flares show both a different structure and peak energy flux 
(
$9.6 \pm1.8 \times 10^{-10}$ \ergcms and 
$7.1 \pm1.3 \times 10^{-10}$ \ergcms, 
respectively in 2010 and 2014). While  the tail of the 2010 flare the flux decayed exponentially, in 2014 it persisted at a high level. The interpretation of these differences as well as of the flare themselves is subject of debate.}

\FullConference{Swift: 10 Years of Discovery,\\
		2-5 December 2014\\
		La Sapienza University, Rome, Italy }

\begin{document}

\section{Introduction}

\psrb/LS 2883 is one of the few binaries systems belonging to the class of the "$\gamma$-ray binaries". This class is a subsample of the High Mass X-ray Binaries (HMXB) with the peculiar characteristic that the $\gamma$-ray emission dominates the radiative spectral energy distributions \cite{Dubus2013}.  
At present the confirmed $\gamma$-ray binaries are:
 \lsi\ \cite{2006Sci...312.1771A, 2008ApJ...679.1427A, 2009ApJ...701L.123A,2012ApJ...749...54H},
 \ls\ \cite{2005Sci...309..746A,2009ApJ...706L..56A,2012ApJ...749...54H},
 PSR B1259$-$63 \cite{2005A&A...442....1A,B1259-Abdo2011,B1259-Tam2011},
 \hess\  \cite{hess_discovery, hinton2009, Bongiorno2011},
 1FGL J1018.6$-$5856 \cite{1FGLbinary_science}.

Among them, \psrb\ is the only one whose compact object is known to be a radio pulsar. 
It is a young pulsar with spinning period $P=47.76$ ms \cite{Johnston1994}, that moves in a very eccentric orbit ($e\sim8.7$) around the massive star LS 2883, with orbital period of 1236.7 days ($\sim$3.4 years) \cite{Shannon2014}.
LS 2883 is a massive Be star ($M \sim 30 M_{\odot}$) with an equatorial disk inclined with respect to
the orbital plane, such that the pulsar crosses the disk just before and just after each periastron passage \cite{Wex1998}.

 The powerful pulsar wind generated by the pulsar interacts with the stellar disk during the passages through it, giving rise to a broadband non-thermal unpulsed emission in radio (e.g. \cite{Johnston2005}), X-rays (e.g. \cite{Chernyakova2009}), and TeV $\gamma$-rays \cite{Aharonian2005, Aharonian2009}.

 In the GeV energy band, the first periastron passage observed by Fermi-LAT occurred in December 2010 \cite{B1259-Abdo2011, B1259-Tam2011}.
 The behavior of the source in this energy range was surprisingly different from the other wavelengths.  
 Before and during the periastron passage, the LAT detected just a weak emission above 100 MeV, while $\sim$30 days after the periastron a very bright flare was detected, with a flux about 10 times the
pre-periastron value. The flares continued until about three months after the periastron passage.

The radiation mechanisms of the broadband emission from \psrb/LS 2883, as
well as the GeV flare is still under debate (e.g. \cite{Khangulyan, Kong2012, DubusCerutti2013, Chernyakova2014}). 
 
 After its 3.4-year orbit period, \psrb\ approached periastron again on 2014 May 4. Fermi-LAT has monitored this further passage \cite{Tam2014, Caliandro2015, Tam2014, atel6204, atel6216, atel6225, atel6231}.
 In this proceeding, Fermi-LAT analysis results of \psrb\ over the last periastron passage are presented, together with the comparison and a new analysis of the periastron passage occurred from 2010.

\section{Periastron passages during Fermi-LAT activity}

Since the launch of the Fermi satellite \psrb\ undergo two periastron passages, on 2010, and more recently on 2014. Precisely, as derived from the most recent measurement of orbital parameters \cite{Shannon2014}, the 2010 and 2014 periastron time are MJD 55544.693781 (2010 December 14 16:39:02.678 ) and MJD 56781.418307 (2014 May 4 10:02:21.725). 
We analyzed Fermi-LAT data since 2 months prior to each periastron up to three months after. 
The analysis is performed selecting from the Pass 7 reprocessed data SOURCE class events which fall in a circular region of $10^{\circ}$ radius centered on \psrb, and within an energy range of 0.1-100 GeV. To reject contaminating gamma rays from the Earth's limb, we selected events with zenith angle smaller than $100^{\circ}$.
The light curves presented here are calculated performing a binned maximum likelihood fit in each time bin using the Science Tool {\em gtlike}.
The spectral-spatial model constructed to perform the likelihood analysis
includes Galactic and isotropic diffuse emission components as well as known $\gamma$-ray
sources within $15^{\circ}$ of \psrb\ based on the third Fermi-LAT source catalog \cite{3FGL}. The
spectral parameters were fixed to the catalog values, except for the sources within $3^{\circ}$ of \psrb. For these latter sources, the flux normalization was left free. 
\psrb\ itself was modeled as a single power-law with all spectral parameters allowed to vary.
For the time bins with low detection significance (Test Statistic value TS < 9), we placed a 95\% upper limit on the \psrb\ photon flux above 100 MeV, as evaluated with Helene's method \cite{Helene1983}. 

\begin{figure}
\center
\begin{minipage}{.37\textwidth}
  \flushleft
  \includegraphics[width=.99\linewidth, trim=0.6cm 0 0 0]{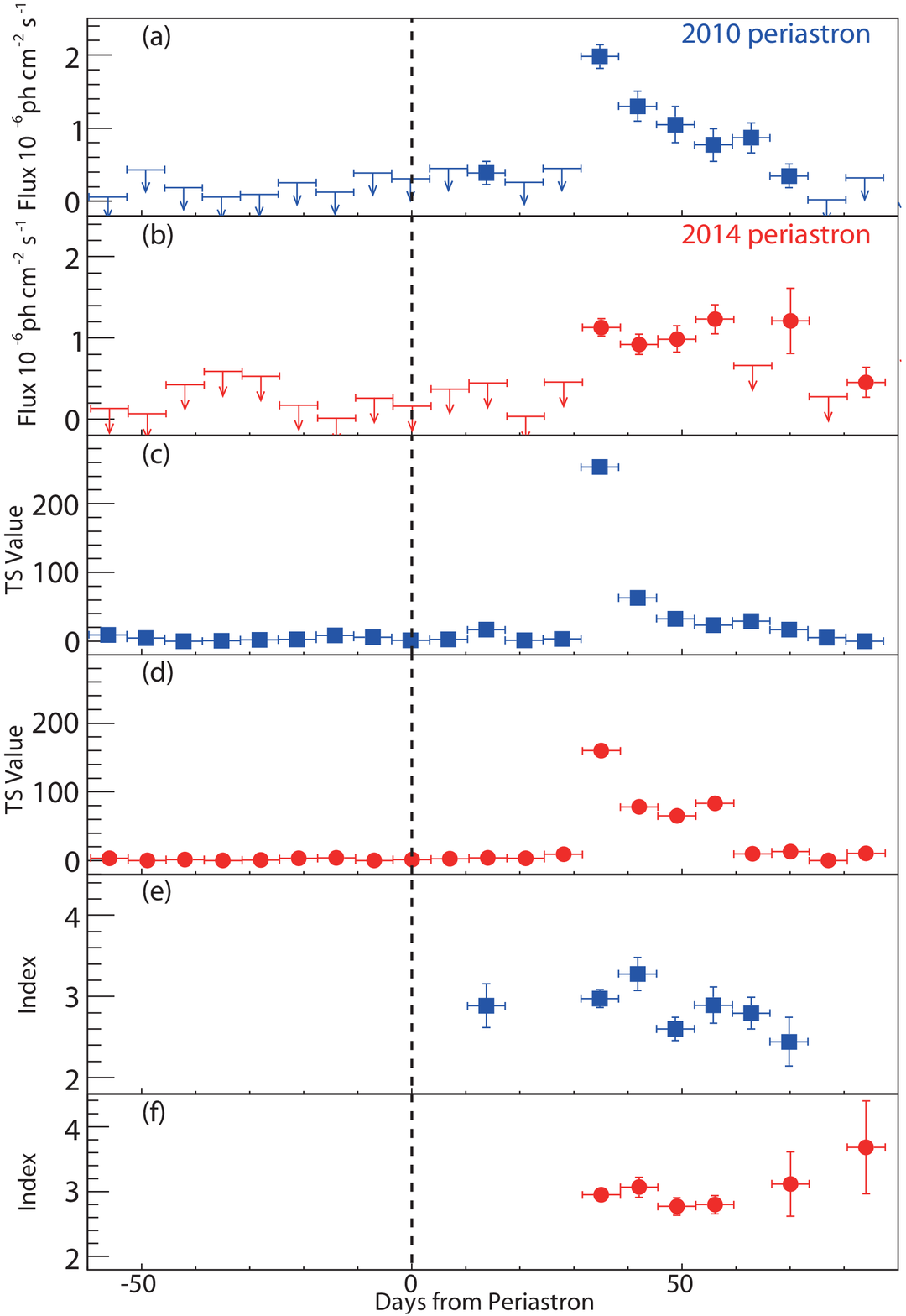}
\end{minipage}%
\hspace{0.5cm}
\begin{minipage}{.58\textwidth}
  \flushright
  \includegraphics[width=.99\linewidth, trim=0 0 2.0cm 0]{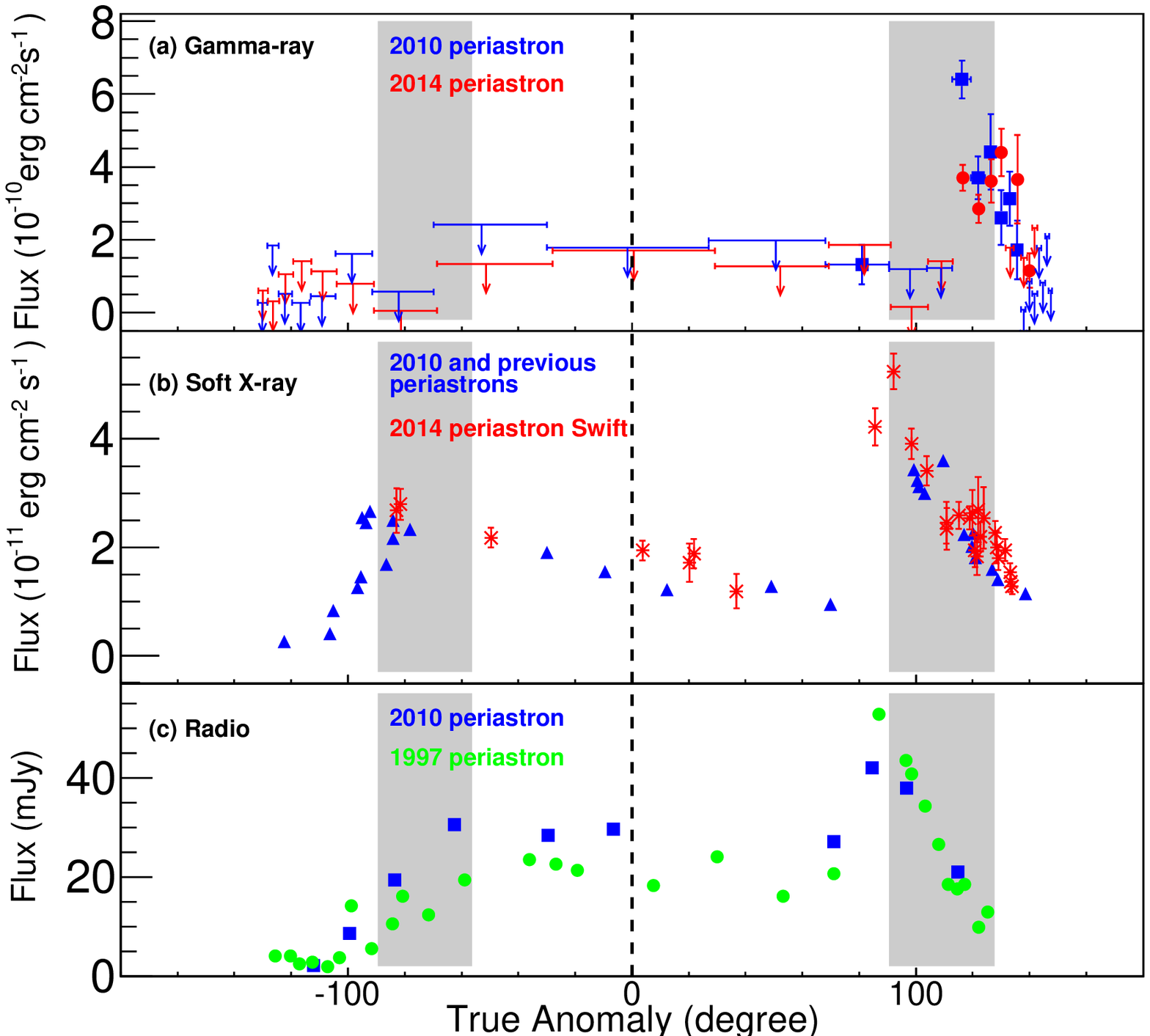}
\end{minipage}%
\caption{{\bf Preliminary.} {\em Left.} Weekly $\gamma$-ray flux for $E>100$ MeV (a \& b), TS
value (c \& d), and photon index (e \& f) of \psrb\ during the 2010 (blue)
and 2014 (red) periastron passages. The dashed black line shows the time of periastron.
{\em Right.} (a) Weekly $\gamma$-ray flux of 2010 (blue) and 2014 (red) periastron passage in energy units. 
(b) X-ray fluxes from 2014 Swift (red) and previous periastron passages (blue, 2004, 2007 and 2010 periastra
from \cite{Chernyakova2009}). (c) Radio (2.4 GHz) flux densities for the 2010 (blue) and
1997 periastron passages (green, \cite{B1259-Abdo2011}). X axis is in true anomaly. The dashed
black line shows the time of periastron. Shaded area corresponds to the Be circumstellar
disk position proposed in \cite{Chernyakova2006}.
}
\label{WeeklyLC}
\end{figure}

In Figure \ref{WeeklyLC} (left panel) are shown the weekly light curves of the 2010 (blue), and 2014 (red) periastron passages.
In both of them a bright flaring activity below 1 GeV was detected starting $\sim$30 days after the periastron.

It is important to highlight that on 2014 Fermi-LAT performed a Target of Opportunity (ToO) observation of \psrb\ starting 27 days after the periastron passage, and lasting 19 days. Thus, the prompt of the gamma-ray activity was observed with an exposure of \psrb\ increased by a factor of $\sim$2 compared to normal observations.    

The analysis performed here showed consistent results with those reported in \cite{B1259-Abdo2011}, in which the analysis was carried out with Pass 6 data.

\section{Similarities and differences between 2014 and 2010 flares}

The onset of both 2010 and 2014 flares is after $\sim$30 days from the periastron. 
They have a similar duration of $\sim$40 d.
The flaring is characterized by a sudden high flux increase followed by a tail, which on 2010 exponentially decrease, while in 2014 is flatter, as shown in Figure \ref{WeeklyLC} left panel.

\begin{figure}
\center
\includegraphics[width=0.99\textwidth]{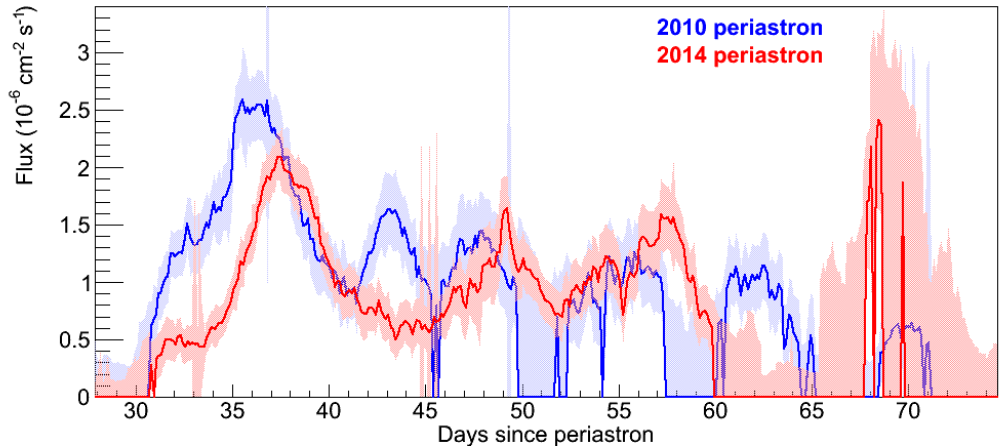}
\caption{{\bf Preliminary.} Flux smoothed light curves of the 2010 and 2014 flaring activities
after periastron (blue and red, respectively) produced with a sliding window. The shaded
areas demonstrate the statistical error of the flux. The smoothed light curve points are
represented as null flux when the TS value is below 9, with only shaded areas showing the
upper limits of 95\% confidence.
}
\label{Smooth}
\end{figure}

In order to study similarities and differences between the 2014 and 2010 flare
 profiles, their smoothed light curves are plotted in Figure \ref{Smooth}. The smoothed light curves
 were produced using a sliding window technique. We chose a time window of 3 days moving
 forward in time with steps of 3 hours. For each step, a binned likelihood analysis was
 performed on the time window. The spectral index of \psrb\ was allowed to vary
 between the values 2.0 and 3.5.
 The shaded areas in Figure \ref{Smooth} around the solid lines correspond to the
 statistical error of the flux calculation. When the TS value is low (TS < 9), the smoothed
 light curve points are represented as null flux with shaded areas showing the upper limits of
 95\% confidence level.
 
 After the onset, the 2010 periastron smoothed flux steeply increases up to its maximum at $\sim$36
 days. However, in 2014 the increase is delayed by forming a short plateau until $\sim$34 days,
 after which the flux rises up to the peak (at $\sim$38 days). 

 The peak flux of 2014 periastron is lower than the 2010 periastron.
Indeed, the 2014 periastron highest day-average flux was $7.1 \pm1.3 \times 10^{-10}$ \ergcms, corresponding to $\sim$50\% 
of the pulsar spin-down luminosity, $8.3 \times 10^{35}$ erg s$^{-1}$ \cite{Johnston1994}.
In contrast, the highest day-average flux of 2010 periastron was $9.6 \pm1.8 \times 10^{-10}$ \ergcms, corresponding to $\sim$70\% of the pulsar spin-down luminosity.

 The main peak of the 2014 profile is delayed by $\sim$2 days with respect to the 2010 flare.
  This delay is consistent with the result of the cross-correlation we calculated for the daily light curves of the 2010 and 2014 flaring activity (The daily light curves are shown in \cite{Caliandro2015}).
   
   In 2010 the main peak is followed
 by a marked second peak (at $\sim$43 days), which in the 2014 light curve is substituted by
 a smooth valley. 
 
 Finally, the 2014 flaring activity ends with a possible peak at $\sim$68 days,
 which is consistent with the last high flux bin of the weekly light curve (see Figure \ref{WeeklyLC}). 
 However, the TS value of this bin is very low (TS$<20$). We have also noticed that the exposure of \psrb\ reaches its minimum exactly in correspondence of this final peak. It is $\sim$5 times less than the exposure of the main peak.
 Therefore, this final peak can be just an artifact due to the very low exposure.

\section{Multi-Wavelength context}

During its 2014 periastron \psrb\ was monitored also by Swift/XRT in Photon Counting (PC) and Window Timing (WT) modes. We analyzed the Swift/XRT 0.1-10 keV data. A detailed description of the analysis is given in \cite{Caliandro2015}.

In figure \ref{WeeklyLC} (right panel) the Fermi-LAT light curves are compared with  
 X-ray measurements of \psrb\ during 2014 and previous periastra, and radio measurements during 2010 and 1997 periastra.
 
 The Radio and X-ray flux profiles show an enhancement either before and after the periastron, 
 which is expected when the
 pulsar crosses the companion's circumstellar disk twice each orbit \cite{Cominsky, Chernyakova2006, Chernyakova2009, B1259-Abdo2011}. In Figure \ref{WeeklyLC} (right panel) the position of the Be
 circumstellar disk is shown with shaded areas as proposed in \cite{Chernyakova2006}.
 
In contrast, the $\gamma$-ray flaring activity was detected only after the periastron.
Although the 2010 and 2014 flares are within the second crossings of the Be circumstellar disk, they are not in line with X-ray and radio activities. Indeed, the prompt of the $\gamma$-ray flares are delayed respect to the other wavelengths by $\sim$30 true anomaly degree, corresponding to $\sim$15 days.

\section{Summary}

The observations of 2010 and 2014 periastron passages of \psrb\ with Fermi-LAT has demonstrated the recurrent flaring behavior of the gamma-ray emission after the periastron passage. We have shown that the two passages manifest some striking similarities. At the same time, there are also certain differences which may be due to inhomogeneities in the shape, density, or extent in the circumstellar disk of the Be star. The two flares were characterized by a high efficiency of conversion from pulsar spin-down power into gamma-ray emissions. 
	Though we have now observed two similar GeV flares from \psrb, the origin of these events is still unclear. To explain it authors have considered up scattering of the circumstellar disk IR photons by the unshocked pulsar wind electrons \cite{Khangulyan}, or Doppler boosted synchrotron emission in the bow-shock tail \cite{Kong2012}, or also up-scattering of X-ray photons from the surrounding pulsar wind nebula \cite{DubusCerutti2013}.

\section{Acknowledgements}
The \textit{Fermi}-LAT Collaboration acknowledges support for LAT development, operation and data analysis from NASA and DOE (United States), CEA/Irfu and IN2P3/CNRS (France), ASI and INFN (Italy), MEXT, KEK, and JAXA (Japan), and the K.A.~Wallenberg Foundation, the Swedish Research Council and the National Space Board (Sweden). Science analysis support in the operations phase from INAF (Italy) and CNES (France) is also gratefully acknowledged.

\end{document}